\documentclass[sigplan,nonacm]{acmart}
\usepackage{tabularx}

\usepackage{caption}
\usepackage{subfig}
\usepackage{multirow}
\usepackage{booktabs} 
\usepackage{array}    
\usepackage{tcolorbox}

\begin{document}

\title{Reducing the Cost of Dropout in Flash-Attention by Hiding RNG with GEMM}

\author{Haiyue Ma}
\authornote{The work was done while the author was an intern at NVIDIA.}
\affiliation{%
  \institution{Princeton University, NVIDIA}
  \country{USA}
}
\author{Jian Liu}
\affiliation{%
  \institution{NVIDIA}
  \country{USA}
}
\author{Ronny Krashinsky}
\affiliation{%
  \institution{NVIDIA}
  \country{USA}
}


\begin{abstract}
Dropout, a network operator, when enabled is likely to dramatically impact the performance of Flash-Attention, which in turn increases the end-to-end training time of Large-Language-Models (LLMs). The main contributor to such performance degradation is the Random Number Generation (RNG) phase. The state-of-the-art optimization is to fuse RNG into the Flash-Attention kernel. However, while RNG and Attention do not compete on compute or memory resources, they are bounded on the same lower-level architecture bottlenecks. Fusion can hardly hide RNG latency within the Attention kernel. 

We propose overlapping RNG with previous GEMM layers in the network to hide RNG latency and improve end-to-end performance. RNG and GEMM have distinct resource requirements and hardware bottlenecks, so they can run together without compromising each other's performance. We propose a fine-grained analytical performance model that analyzes low-level architecture resource utilization to evaluate RNG-GEMM overlapping performance benefits. This model, cross-validated by silicon results, shows 1.26x speedup for overlapping RNG and GEMM layers over a sequential implementation on one Transformer Block (one LLM layer including multi-head attention and feed-forward layers), and 1.22x over state-of-the-art fusion implementation, for Llama3 on GH100 GPUs with FP8 precision. Because the kernel patterns are regular, the findings of the shared bottlenecks, as well as the achievable performance benefits, can be generalized to different model architectures, software implementations and hardware configurations. 

\end{abstract}

\maketitle

\section{Introduction}
Large Language Models (LLMs) have become important targets for performance optimization due to their ever-increasing workload sizes and corresponding runtime demands. Full-scaled training for big models requires several months and thousands of GPUs~\cite{openai2023gpt4}\cite{gpt4_trainingtime}. Attention dropout~\cite{srivastava2014dropout}\cite{zehui2019dropattention} is an optional technique that drops out elements after the Softmax operation in Attention. Dropout is applied in commonly used models such as Llama~\cite{llama_git}\cite{touvron2023llama} because it can make the model focus on relevant features and improve training accuracy~\cite{attention_dropout}\cite{attention_dropout2}~\cite{xue2024repeat}. However, enabling dropout is costly, which doubles the processing time of the Attention layer with state-of-the-art implementations like Flash-Attention~\cite{dao2022flashattention}\cite{dao2023flashattention}\cite{shah2024flashattention}, and in turn increases the end-to-end training time by 1.3x to 1.7x. Optimizing the runtime of dropout can significantly improve the performance of LLM training.

The runtime of dropout is dominated by the Random Number Generator (RNG)~\cite{philox}, which generates random numbers to determine which elements within the intermediate matrix of the Attention layer to drop. The size of this matrix, and consequently the runtime of RNG, scales quadratically with the sequence length. The industry trend towards ever larger sequence lengths further exacerbates the RNG latency.

\begin{figure}[t]
\centering
\includegraphics[width=0.5\textwidth]{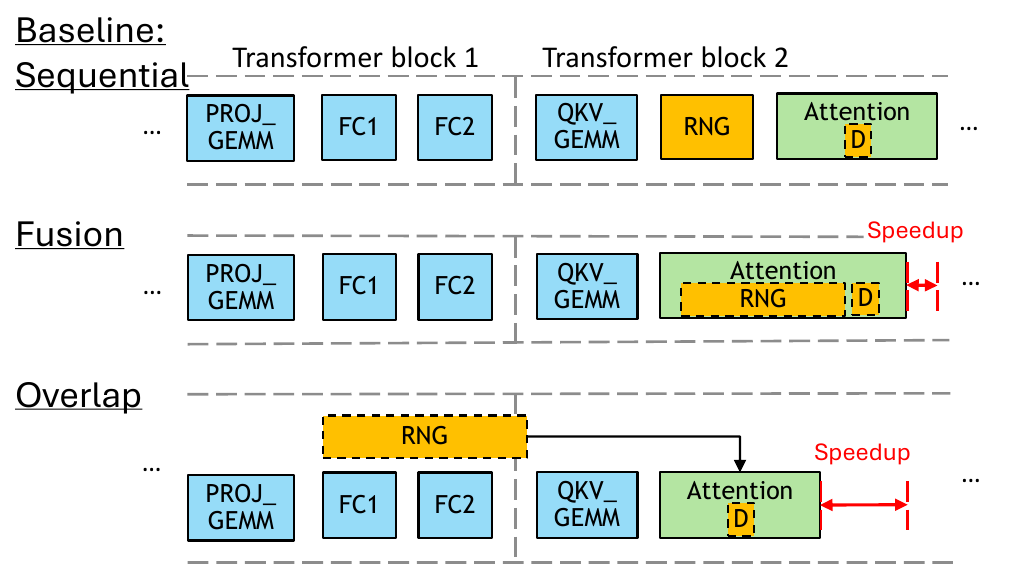}
\caption{Attention has a dependency on RNG and Dropout (``D''). State-of-the-art optimization is kernel fusion, which results in hardware bottleneck conflicts. We propose overlapping RNG with GEMM layers to achieve significant speedup.}
\label{fig:overview}
\end{figure}

A starter implementation is to run RNG sequentially before Attention, and store the random numbers to generate the masks for dropping out elements (``D'') in Flash-Attention, as shown in Figure~\ref{fig:overview}. To improve performance, state-of-the-art solution fuses dropout (including RNG) into the Flash-Attention kernel. A common but misleading assumption is that Flash-Attention in LLM training is dominated by matrix-matrix multiplication (MMA) math and thus limited by compute capability. This assumption rules out the possibility that Attention and RNG share the same hardware bottlenecks, because RNG has no MMA operations. However, with a fine-grained analysis with more lower-level architecture bottlenecks on the kernel code, we observe that Attention is actually limited by non-compute hardware resources: the Register File bandwidth, and the Issue Stage bandwidth. Because RNG is also bottlenecked by the Issue Stage, its latency is almost fully exposed when fused with Attention. Silicon measurements show that only 10-20\% of RNG runtime can be hidden within the Attention kernel.

In this paper, we propose overlapping RNG with other matrix multiplication (GEMM) layers to hide RNG runtime and improve end-to-end performance. 
RNG and GEMM are ideal targets for overlapping because they do not depend on each other and have distinct resource requirements and hardware bottlenecks. GEMM is mostly bounded by MMA math and L2 bandwidth, whereas RNG mostly by the Issue Stage and ALU operations. While GEMM has a heavy consumption on the Register Files (RF) and Shared Memory (SMEM) within each Streaming Multiprocessor (SM), RNG uses minimal RF and SMEM, allowing GEMM to run in parallel with RNG while obtaining near-optimal performance.

We built a fine-grained analytical model to evaluate the performance for different kernel implementations: sequential execution, RNG-Attention fusion, and RNG-GEMM overlapping. This model analyzes hardware bottlenecks beyond compute and memory, and takes into account lower-level hardware configurations such as Register File, Issue Stage, and instruction execution pipelines. It takes in GPU hardware configurations and the model parameters, and outputs the runtime for kernels based on the workload's primary bottleneck. Our model is validated by silicon results running with FP8 precision~\cite{fp8_model} on H100 HBM3 80GB GPUs~\cite{h100_whitepaper}. 

Results demonstrate great potential when overlapping RNG with the GEMM layers between the previous and the current Attention layer: for Llama3~\cite{grattafiori2024llama}, overlapping achieves 1.26x speedup compared to baseline sequential kernels, and 1.22x when compared to state-of-the-art fusion methodology; similar speedup is obtained for GPT-4~\cite{openai2023gpt4}. Since the identified bottlenecks for the analyzed kernels remain the same for regular LLM training workloads with dense, non-narrow matrix shapes, the performance benefits can be generalized.




Using the analytical model, we derive implications of RNG-GEMM overlapping technique under different hardware and software settings. As a general trend, future generations of GPUs advance in computing power (Tensor Cores), but are likely to have similar architecture designs for non-Tensor units such as Register File, Issue Stage and ALU pipelines. On these future GPUs, non-Tensor resource bottlenecks are even more severe, and overlapping RNG and GEMM is more favorable. 

The analytical model is widely applicable with configurable inputs for model architecture (such as Sequence Length, Batch Size, Embedding Dimension), software implementations (such as tile size, kernel ordering), and hardware configurations (such as compute capability, memory bandwidth, Register File, Issue Stage and execution pipelines). 
Further, the model can be easily adapted to an universal overlapping/fusion scheduler evaluator that works for all operators.


In this paper, we make the following contributions: 
\begin{itemize}
    \item We proposed overlapping the key component of dropout - RNG - with GEMM kernels to hide RNG latency, leading to substantial speedup across multiple trending LLM models compared to state-of-the-art fusion technique.
    \item We developed an analytical performance model based on fine-grained hardware bottlenecks beyond compute, communication and memory. This model accurately derives performance comparisons between sequential kernels, fused kernels and overlapping kernels. It is cross-validated by results obtained from real silicon implementations.
    \item We explored the broader implications of the overlapping mechanism for various model architectures, future GPU hardware designs, and different software implementations. 
\end{itemize}
\section{Background}

\begin{figure}[t]
\centering
\includegraphics[width=0.5\textwidth]{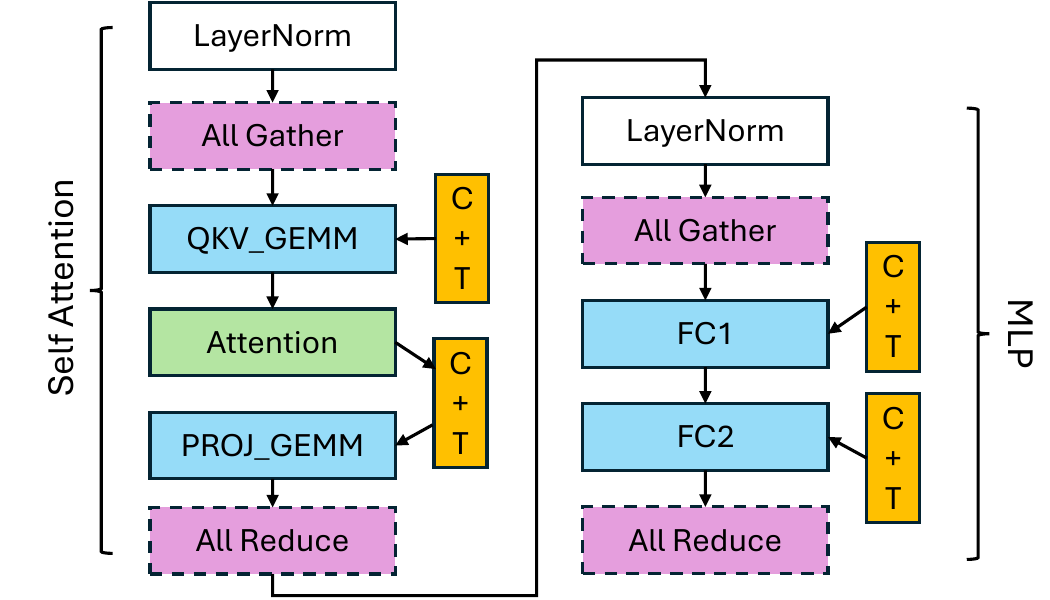}
\caption{Network Architecture of one Transformer Block, the basic building block of LLM. The four GEMM and the Attention layers dominate the runtime, and the communication latency is typically hidden within the GEMM layers.}
\label{fig:llm_network}
\end{figure}

\subsection{LLM Network Architecture}
LLM networks typically begin with an embedding layer, conclude with a decoding layer, and iteratively call Transformer Blocks in between. Figure~\ref{fig:llm_network} shows the network architecture of one Transformer Block in the forward path. The General Matrix Multiply (GEMM) layers in blue and the Attention layer in green contribute to the majority of the compute time. The purple dashed layers represent communication layers, present only in multi-GPU systems. The white LayerNorm layers perform element-wise operations, and the orange C+T layers handle Conversion (when precision conversion to FP8 is necessary) and Transpose operations for GEMM inputs. These last two types of layers typically require minimal runtime and are omitted from our runtime analysis.

In this work, we begin our analysis on single GPU without communication, and focus on the GEMM and Attention layers. We discuss multi-GPU scenarios in Section~\ref{sec:discussion} and draw similar conclusions as in single GPU. With multi-GPU, different parallelism mechanisms~\cite{shoeybi2019megatron} are applied which evenly distribute workload onto different GPUs. This does not impact savings of overlapping since the ratio of each kernel's runtime remains the same. In highly optimized implementations, communication layers are often overlapped with their producer layers to minimize latency overhead, therefore the end-to-end savings brought by overlapping still apply.

\subsection{Dropout}
\begin{figure}[t]
\centering
\includegraphics[width=0.45\textwidth]{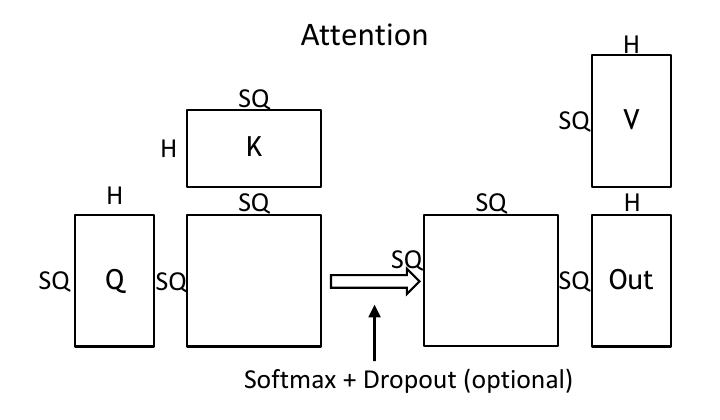}
\caption{Dropout is applied to the intermediate results in Flash-Attention, after the Softmax operation.}
\label{fig:dropout_location}
\end{figure}

Dropout~\cite{srivastava2014dropout}\cite{wan2013regularization} is a technique designed to prevent overfitting by randomly setting a small fraction of the elements within a network to zero. It has been proposed to regularize neuron networks in general. In the context of this paper, we refer specifically to Attention dropout~\cite{attention_dropout}\cite{attention_dropout2}, where the dropout operation is applied to the intermediate outputs of the Attention layer following the Softmax operation, as depicted in Figure~\ref{fig:dropout_location}.

While optional, dropout can substantially improve training accuracy: it prevents the model from relying heavily on certain features, thus making the model only focus on relevant features to prevent overfitting and underfitting~\cite{xue2024repeat}\cite{liu2023dropout}. Dropout is used in training widely adapted LLM networks such as Llama~\cite{llama_git}\cite{touvron2023llama}. However, its huge runtime slowdown has limited its application. With our dropout optimizations, we can enable its usage in future applications.

\subsection{RNG Implementation: Philox}
The computational demand of dropout mostly comes from the Random Number Generator (RNG) used to determine which elements to zero out. Multiple possible RNG implementations exist, and our discussion centers on Philox~\cite{philox}~\cite{philox_numpy}. Philox is a counter-based pseudorandom number generator (PRNG) that relies on wide multiplies and iterates over previous states to generate new ones. Philox is well-suited for GPU execution, as its operations can be efficiently parallelized. Implementations of Philox are available in NVIDIA's cuRAND library~\cite{curand} and TensorFlow~\cite{tensorflow_rng}. Our analysis primarily focuses on a seven-iteration implementation (Philox 7), though we also consider more resource-efficient versions (Philox 5 and 3) in Section~\ref{sec:diff_rng}.

\section{Analytical Model Methodology}

In this section, we explain the construction of our fine-grained analytical performance model. We discuss the hardware resources and limiters, the software implementations of LLM networks, and the derivation of runtime of individual kernels, kernel fusion and kernel overlapping.


\subsection{Hardware Limiters}
We identified several hardware limiters critical for calculating kernel runtime in our performance model. For each kernel, we estimate its runtime assuming each limiter as a potential bottleneck and determine the final runtime by selecting the maximum runtime among all considered limiters. 

Our model is at tile-level granularity because LLM implementations are usually tiled. For MatMul kernels, we first model one tile's runtime, then derive the total runtime by multiplying with the total number of tiles. For non-MatMul operators such as stand-alone RNG, we directly model the entire kernel's runtime based on the total number of operations and the primary bottlenecks. 

\begin{table*}[t]
\caption{The hardware limiters included in our analytical model.}
\centering
\begin{tabularx}{0.99\textwidth}{|>{\hsize=0.3\hsize}X|>{\hsize=0.65\hsize}X|>{\hsize=0.35\hsize}X|>{\hsize=0.45\hsize}X|>{\hsize=0.25\hsize}X|}
\hline
Limiter & Hardware Resources & \multicolumn{3}{c|}{Total Operations for Each Kernel}\\ \hline
 & & GEMM ($A_{m,k}B_{k,n}$) & Attention ($Softmax (Q_{m,n}K_{m,n}^T)V_{m,n}$) & RNG ($m*m$ Dimension) \\
\hline\hline
MMA Math & Tensor Core \#FLOP per second & \#MMA ops: $m*n*k$  & Same as GEMM for two MatMuls: $m*n*m*2$ & / (No MatMul) \\\hline
HBM Read BW & Read bandwidth of the main memory & \#inputs: $m*k+n*k$ & Same as GEMM for two Matmuls &  / (No memory reads)\\\hline
L2 Read BW & Read bandwidth between L2 cache and SM cores & $\#inputs*\#SMs$: $k(m*num\_n\_tiles + n*num\_m\_tiles)$ & Same as GEMM for two Matmuls & / (No memory reads) \\\hline
RF Read BW & Each SM's Reg File read bandwidth & \multicolumn{3}{c|}{ \#Types of instructions * \#RF reads in each type} \\\hline
Issue Stage & Pipeline for issuing instructions & \multicolumn{3}{c|}{Number of instructions} \\\hline
ALU Pipe & ALU instruction pipe & \multicolumn{3}{c|}{Number of ALU instructions in kernel}  \\\hline
MUFU Pipe & MUlti-FUnction instruction pipe & \multicolumn{3}{c|}{Number of multi-function unit instructions}  \\\hline
FMA Pipe & Fused Multiply-Add instruction pipe & \multicolumn{3}{c|}{Number of FMA instructions} \\\hline
\end{tabularx}
\label{tab:limiter}
\end{table*}

The hardware limiters that we chose to investigate in our model are listed in Table~\ref{tab:limiter}. These limiters are all configurable in order to model different hardware designs. For each row, runtime is calculated by the number of operations divided by the hardware resources available. Total operations for MatMuls are obtained from the theoretical number of mul-add operations based on matrix dimensions. To further improve model accuracy, with the access to production-ready, highly optimized kernel code, we obtain non-MatMul kernels' number of operations by counting the number of instructions from the production kernel's assembly code (SASS). These numbers can also be derived theoretically without the access to kernel assembly code. 

\begin{itemize}
    \item MMA Math: Tensor Core TFLOPS obtained from GPU datasheet. GEMM and Attention kernels during LLM training perform dense matrix multiplications and are mostly bounded on MMA Math. Runtime is calculated by the number of MMA operations ($m*n*k$) divided by TFLOPS. RNG kernel does not contain MMA operations. 
    \item Main-memory Read BW: The memory (HBM in high-end datacenter GPUs) read bandwith in Bytes per second. Runtime is calculated by total number of memory reads divided by the memory bandwidth for GEMM and Attention layers' matrix multiplications. RNG kernels only generate random numbers and there is no memory read. 
    \item L2 Read BW: Bandwidth for L2 cache reads. Cache reads for GEMM and Attention kernels are calculated as the size of each input Matrix multiplied by the number of SMs that access a certain element in the matrix. Since tiles are being computed on different SMs, each element needs to be read by multiple SMs. RNG does not read from memory nor does it have L2 reads. 
    \item RF Read BW: The register file read bandwidth inside each SM core. For all kernels, this is calculated by going through each type of the frequent instructions in the kernel, and obtaining the number of registers that each instruction type uses. For example, FMA (Fused Multiply-Add) instruction has three input registers. While GEMM typically uses a lot of registers due to its heavy matrix multiplication operations, this bottleneck is secondary compared to compute (MMA Math). However, Flash-Attention has softmax that heavily uses FMA and other operations with high demand of register file read bandwidth. The softmax demand on top of the $(QK^T)V$ matrix multiplication makes register read bandwidth one of the most critical hardware bottlenecks for Flash-Attention. 
    \item Issue Stage: Typically one instruction is issued each cycle. Runtime is obtained by analyzing the number of total instructions in each kernel. RNG and Attention kernel (with softmax) both have a lot of instructions and are typically bounded by the issue stage. 
    \item ALU/MUFU/FMA Pipe: Number of ALU/MUFU/FMA instructions in the kernel divided by each function unit's throughput. RNG kernel is heavily bounded on ALU pipe, while the Attention kernel has a lot of MUFU/FMA instructions. 
\end{itemize}

We focus on read bandwidths because typically applications will have more reads than writes that go through any unit in the memory hierarchy. In practice, we also observe that when bandwidth becomes an issue, the real limiter is usually the read bandwidth but not the write. 

\subsection{Software Configurations}

On the software implementation side, the model architecture we used is the vanilla Transformer architecture~\cite{vaswani2017attention}. Specifically, we model the QKV\_GEMM, Flash-Attention (with softmax), PROJ\_GEMM, FC1 and FC2 kernels, which are the main contributors to the runtime of a Transformer Block in LLM training. We do not model the communication layers because in practice, communication latency is shorter than the computation latency and they utilize completely different hardware resources. Communication latency can easily be hidden within the GEMM/Attention layers. 

The configurable inputs to the model from the software implementation side include: 
\begin{itemize}
    \item Batch Size 
    \item Input Sequence Length
    \item Head Dimension
    \item Embedding Dimension (= \#Heads * Head Dimension) 
    \item Tile Size 
\end{itemize}

\subsection{Methodologies to model kernels}
\begin{figure}[t]
  \centering
  \subfloat[Stand-alone GEMM kernel]{\includegraphics[width=0.15\textwidth]{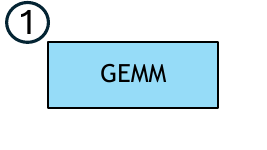}\label{fig:ind_kernel_gemm}}\hfill
  \subfloat[Stand-alone Attention kernel]{\includegraphics[width=0.15\textwidth]{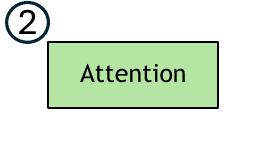}\label{fig:ind_kernel_attention}}\hfill
  \subfloat[Stand-alone RNG kernel]{\includegraphics[width=0.15\textwidth]{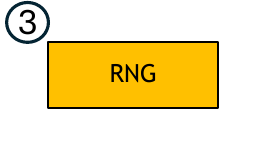}\label{fig:ind_kernel_rng}}\hfill
  \subfloat[Attention with element dropping only]{\includegraphics[width=0.23\textwidth]{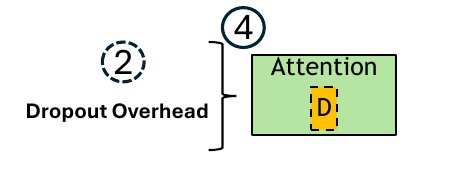}\label{fig:ind_kernel_atten_w_dropout}}\hfill
  \subfloat[RNG and element dropping fused in Attention]{\includegraphics[width=0.23\textwidth]{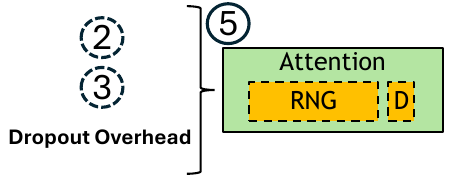}\label{fig:ind_kernel_atten_w_rng}}\hfill
  \subfloat[RNG kernel interfered by GEMM]{\includegraphics[width=0.23\textwidth]{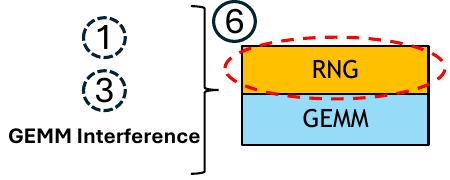}\label{fig:ind_kernel_rng_w_interfere}}\hfill
  \subfloat[GEMM kernel interfered by RNG]{\includegraphics[width=0.23\textwidth]{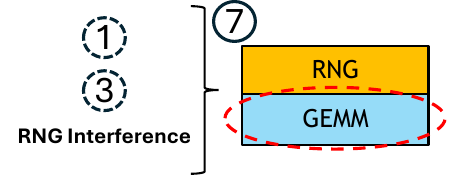}\label{fig:ind_kernel_gemm_w_interfere}}\hfill
  \subfloat[Baseline runtime]{\includegraphics[width=0.15\textwidth]{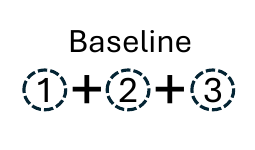}\label{fig:ind_kernel_baseline}}\hfill
  \subfloat[Fusion runtime]{\includegraphics[width=0.15\textwidth]{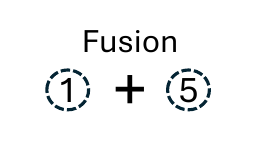}\label{fig:ind_kernel_fusion}}\hfill
  \subfloat[Overlap runtime]{\includegraphics[width=0.15\textwidth]{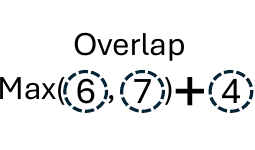}\label{fig:ind_kernel_overlap}}\hfill
  \caption{Theoretical performance model: modeling baseline, fusion and overlap runtime from individual kernel runtime.}
  \label{fig:ind_kernel}
\end{figure}

Figure~\ref{fig:ind_kernel} illustrates the modeling approach for baseline (sequential), fusion and overlap runtime, incrementally constructed with individual kernel performance assessments.

First, we model the performance of each individual kernel - GEMM, Attention, and RNG — by calculating their theoretical runtime based on bounding each by the limiters. For workload sizes within the range of typical networks (GPT~\cite{openai2023gpt4} and Llama~\cite{grattafiori2024llama}), we found that GEMM runtime is bounded by MMA, Attention by RF bandwidth and the Issue stage, and RNG by the Issue stage and ALU pipe.

Next, we integrate additional components to derive runtime estimations for composed kernels. For the Attention kernel with the element dropping step, we derive its runtime by adding the dropping overhead to the standalone Attention kernel runtime (Figure~\ref{fig:ind_kernel_atten_w_dropout}). We also compute the runtime of the fused Attention kernel with RNG by integrating both sets of instructions and identifying the primary limiter, which typically is the Issue Stage, with the ALU pipe and RF bandwidth as close secondary limiters (Figure~\ref{fig:ind_kernel_atten_w_rng}).

We then calculate the runtime of RNG and GEMM separately while the other is running concurrently. RNG runtime is based on the stand-alone RNG, scaled by the GEMM interference overhead. If RNG's runtime exceeds GEMM's, the remaining RNG operations continue at full speed once GEMM completes (Figure~\ref{fig:ind_kernel_rng_w_interfere}). Similarly, we model the GEMM runtime affected by RNG interference (Figure~\ref{fig:ind_kernel_gemm_w_interfere}).

Finally, we determine the baseline runtime from three stand-alone kernels (Figure~\ref{fig:ind_kernel_baseline}), and the fusion runtime from the stand-alone GEMM and the fused Attention-and-RNG kernel runtime (Figure~\ref{fig:ind_kernel_fusion}). The overlap runtime is derived from the maximum runtime of GEMM and RNG with interference, as Attention depends on both GEMM and RNG outputs, plus the standalone Attention runtime with only the element dropping step (Figure~\ref{fig:ind_kernel_overlap}).

In the model, the ratios of overhead and kernel interference are derived from our silicon measurements. Because kernel computation patterns are regular and the bottlenecks are stable, the overhead and interference ratios can be generalized to different implementations. We averaged the numbers across various input workload sizes and model architecture configurations to find the scaling factor that was applied to our model.

\section{Results}

This section presents the performance results of our RNG and GEMM overlapping experiments. We discuss insights derived from our analytical model, and its validation with a silicon implementation. While the analysis presented is based on the Transformer architecture running on H100 GPUs, these conclusions are broadly applicable across various software implementations and hardware configurations. Further analysis of implications on different hardware and software is discussed in Section~\ref{sec:generalization}.

In our study, we assume that GPU resources for LLM training are dedicated, since it's impractical to run other applications in parallel to such a demanding workload. That said, the analytical model is also flexible to model concurrent workloads and shared resources by modifying the configurations to carve out resources for other applications. 

\subsection{Performance Analysis}
We show our experiments on the Transformer network architecture, using a common head dimension of 128, and a batch size of 1. We vary the sequence lengths from 2K to 64K, and embedding dimension from 4K to 16K (number of heads from 32 to 128). We use the tile size 128x128x128 for GEMM and 64x128x128 for RNG and Attention for illustration purposes. These numbers are obtained from the production-level GEMM, Attention, and dropout kernel code, which are chosen for best GPU hardware resources utilization. The exact tile size is not a key factor of the proposal we are making in this paper, and changing the tile sizes in a reasonable range does not impact the final conclusion derived. 

\begin{figure}[t]
\centering
\includegraphics[width=0.5\textwidth]{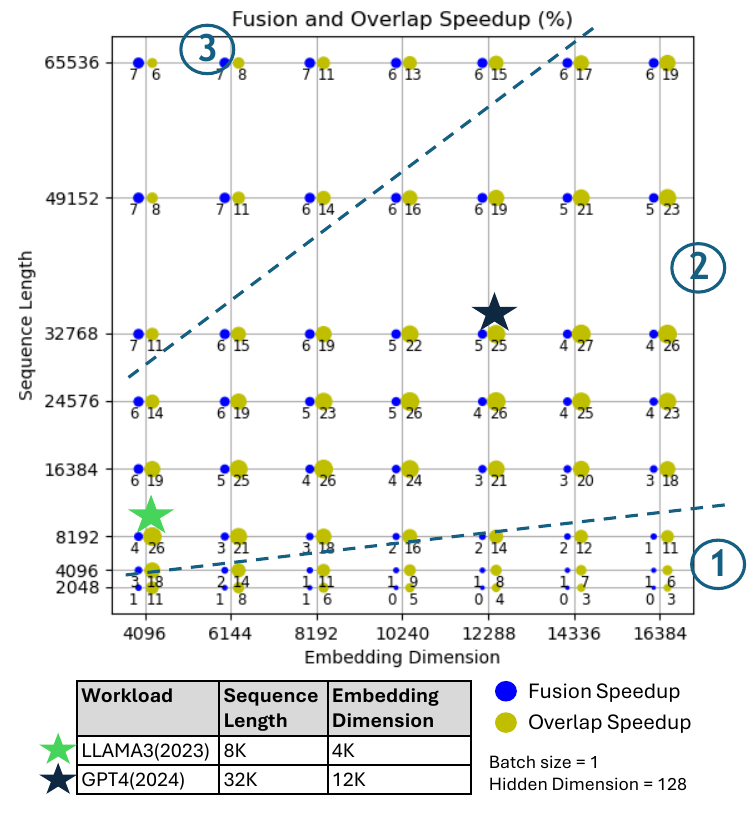}
\caption{Fusion and overlap speedup compared to a baseline sequential kernel implementation for a Transformer Block, obtained from our analytical model.}
\label{fig:4gemm_result_speedup}
\end{figure}

The results demonstrate moderate performance improvements for state-of-the-art RNG-Attention fusion, but significant improvements for proposed RNG-GEMM overlap over the fusion methodology. Specifically, for a Llama3-like workload configuration, we observe 1.26x speedup for the overlap methodology compared to the baseline, and 1.22x on top of the fusion methodology. The performance benefits remain similar as models grow larger: for a later GPT4-like configuration, we observe 1.25x speedup for the overlap methodology compared to the baseline, which is 1.2x on top of fusion. These results suggest that the RNG-GEMM overlap methodology is widely applicable towards multiple workload sizes, and also adapts easily to future generations of workloads. 

The performance benefits change with sequence length and embedding dimension sizes (number of heads * head dimension). To understand these trends, we examined the runtime dependencies of each kernel on these parameters: 

The GEMM layer's runtime depends on the number of multiply-adds, which is $M*N*K$. For each of the four GEMM layers discussed in the paper, the M dimension is proportional to $batch\_size(B) * sequence\_length(SQ)$, and the N and K dimension is proportional to $number\_of\_heads(nH) * hidden\_dimension(dH)$:
\[Runtime(GEMM) = O(B*SQ*dH^2*nH^2)\] 
Similarly, the Attention runtime depends on the number of multiply-adds from the two matrix multiplication:
\[Runtime(Attention) = O(B*nH*dH*SQ^2)\]
The RNG runtime depends on the number of elements in the intermediate layer of Attention:
\[Runtime(RNG) = O(B*nH*SQ^2)\]

\begin{figure}[t]
  \centering
  \subfloat[Individual kernel runtime with different Embedding Dimension: GEMM scales quadratically and Attention/RNG scale linearly.]{\includegraphics[width=0.49\textwidth]{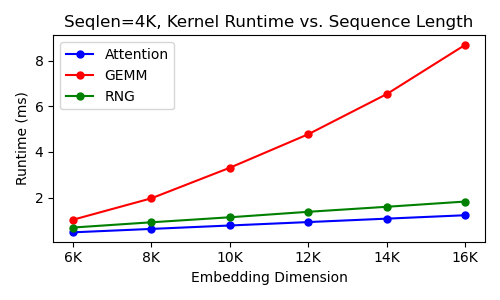}\label{fig:seq4K_heads_v_runtime}}\hfill
  \subfloat[Individual kernel runtime with different Sequence Length: GEMM scales linearly and Attention/RNG scale quadratically.]{\includegraphics[width=0.49\textwidth]{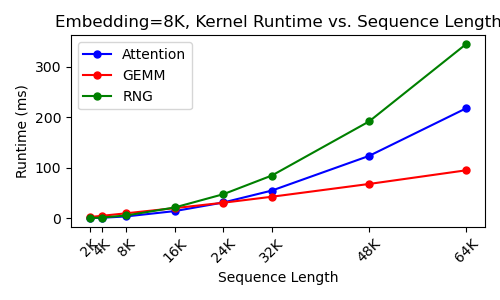}\label{fig:head64_seq_v_runtime}}\hfill
  \caption{Kernel runtime variation with different sequence length and embedding dimension, measured on silicon. }
  \label{fig:kernel_runtime_scale}
\end{figure}

Figure~\ref{fig:kernel_runtime_scale} presents the runtime variation for each kernel across different sequence lengths and number of heads. The data shows that while GEMM runtime scales quadratically with the embedding dimension, the Attention and RNG runtime scales linearly. Conversely, Attention and RNG runtime scales quadratically with sequence length, whereas GEMM runtime scales linearly.

With the overall trends of scaling, we now analyze the numbers shown in Figure~\ref{fig:4gemm_result_speedup}. For RNG-Attention fusion, the absolute runtime saved stays the same regardless of the workload configurations, because Attention and RNG's complexity increases at the same scale. Since we evaluate the runtime for an entire Transformer Block, fusion brings more saving as sequence length increases which also increases the runtime ratio of Attention/RNG. On the other hand, fusion brings less saving as embedding dimension increases where GEMM runtime scales faster than Attention/RNG runtime. Overall, the RNG-Attention fusion speedup is relatively low, which matches our analysis that they have conflicting hardware resource bottlenecks. 

For RNG-GEMM overlap, speedup is correlated with the ratio of sequence length to the embedding dimension, given that the batch size and the hidden dimension don't change (typically they wouldn't). The greatest improvement occurs within region 2 in Figure~\ref{fig:4gemm_result_speedup}, which is the range that common LLM training workloads fall inside.

\begin{figure}[t]
\centering
\includegraphics[width=0.48\textwidth]{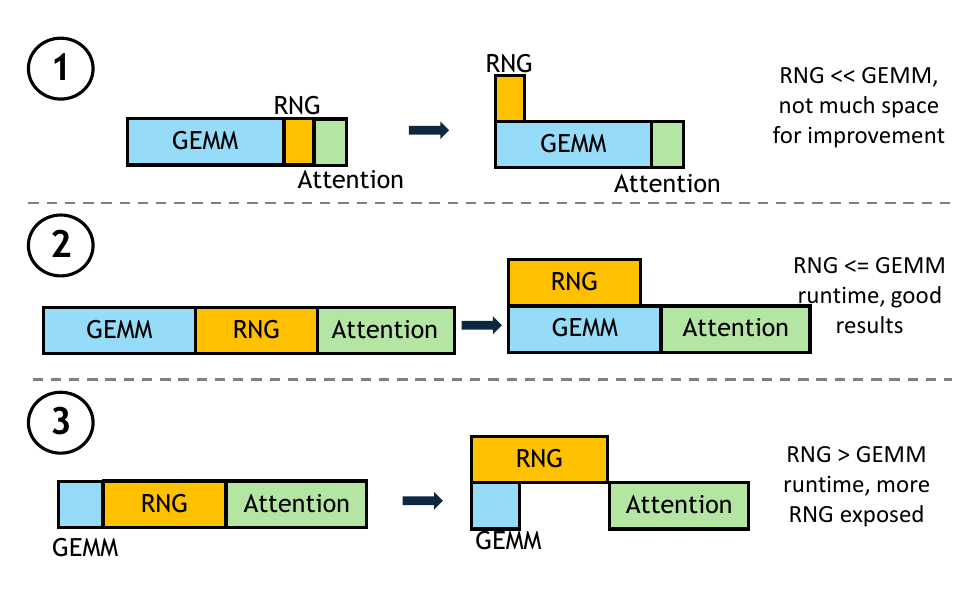}
\caption{Analysis of RNG-GEMM overlapping speedup across three regions in Figure~\ref{fig:4gemm_result_speedup}. Typical LLM training workloads fall inside region 2 with best performance benefits.}
\label{fig:3_region_analysis}
\end{figure}

Figure~\ref{fig:3_region_analysis} illustrates the reasons behind the performance differences of the three regions:
\begin{itemize}
    \item Region 1 (low speedup): Overall runtime is dominated by GEMM layers due to a large number of heads and short sequence lengths. Overlapping offers limited benefits here.
    \item Region 2 (optimal speedup): Balances sequence length and embedding dimension, where RNG runtime is shorter but close to the GEMM runtime. Maximum benefits from hiding RNG latency by overlapping with GEMM.
    \item Region 3 (decreasing speedup): RNG runtime exceeds GEMM because of the long sequence lengths, leading to full exposure of RNG operations post-GEMM completion.
\end{itemize}

As newer generations of LLM training workloads typically grow in both input sequence length and embedding dimension, future, bigger workloads are very likely to stay in region 2 where the RNG-GEMM overlapping technique brings high performance benefits. Skewed workloads that fall in region 1 and 3 are much less likely to appear, and even if that happens, the overlapping technique still outperforms the state-of-the-art RNG-Attention fusion. 


\section{Generalization of Technique}
\label{sec:generalization}

In this section, we explore the implications of the RNG-GEMM overlapping technique with different software and hardware configurations. Specifically, we discuss its performance on future generations of GPUs with higher compute capabilities, and on different implementations of the dropout operator. 

\subsection{Modeling Future Hardware}
\label{sec:diff_hardware}
We use our analytical model to explore how variations in hardware design will influence the efficiency of the overlapping strategy.

A key aspect of GPU hardware evolution is the consistent increase of computational efficiency with each new generation. For example, the NVIDIA Blackwell GPUs~\cite{Blackwell} offer 2x the compute capability as the Hopper generation~\cite{h100_whitepaper}, but the non-Tensor resources (such as register files, issue stage and execution queues) do not see significant improvements. This trend motivates us to model the potential implications of the RNG-GEMM overlapping technique: the conflicting bottlenecks of RNG and Attention will become a more severe issue, which makes the overlapping approach more favorable. 

\begin{figure}[t]
\centering
\includegraphics[width=0.48\textwidth]{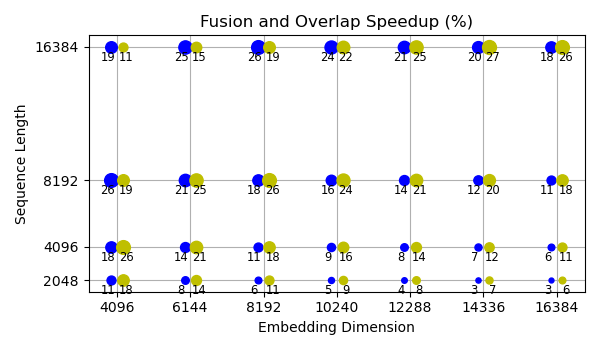}
\caption{RNG-GEMM overlap speedup on a Blackwell B100 GPU model with 2x compute capability (TFLOPs). While GEMM runtime decreases, RNG and Attention runtimes remain constant, leading to a greater proportional speedup.}
\label{fig:diff_hw}
\end{figure}

\begin{figure}[t]
\centering
\includegraphics[width=0.5\textwidth]{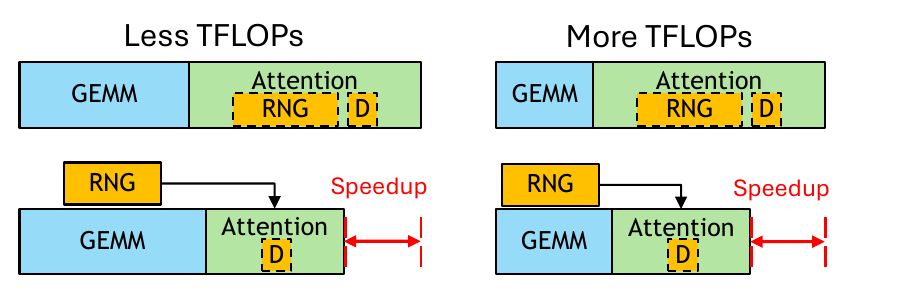}
\caption{Overlap speedup with increased GPU compute capability (TFLOPs). While GEMM runtime decreases, RNG and Attention runtimes remain constant, leading to a greater proportional speedup.}
\label{fig:blackwell_impact_to_overall}
\end{figure}

We evaluated the theoretical performance of our overlapping strategy on a B100 GPU, which offers twice the compute capability of the H100, but other non-Tensor related limiters like the issue pipeline and the ALU pipeline remain unchanged. Computation remains the bottleneck for matrix multiplications, with memory improvements keeping pace with MMA improvements through reduced precision or increased memory bandwidth. 

Figure~\ref{fig:diff_hw} shows our results obtained from modeling workloads with various sequence lengths and embedding dimensions on the B100 chip. Figure~\ref{fig:blackwell_impact_to_overall} illustrates the implications of B100's increased compute capability on our overlapping technique. Since the most significant runtime reduction is observed in the GEMM kernel, this shift makes RNG latency an even more critical bottleneck, increasing its impact on end-to-end network performance. Although the absolute runtime difference between baseline and overlapping remains similar, the relative speedup improves. This difference can be better observed in workloads with shorter sequence lengths; for longer sequences, where RNG and Attention dominate network runtime, overlapping a shorter GEMM could exacerbate problems by fully exposing RNG latency once GEMM computation completes. 

This analysis emphasizes the need for next-generation hardware to consider optimizing traditionally non-Tensor related factors. On the other hand, if hardware optimization on these factors are not possible, we call for future hardware-software co-design to implement kernel overlapping: efficient libraries can be developed to facilitate the overlapping of non-GEMM operations, which typically do not consume extensive computation and memory resources, with GEMM operations.

\begin{figure}[t]
\centering
\includegraphics[width=0.48\textwidth]{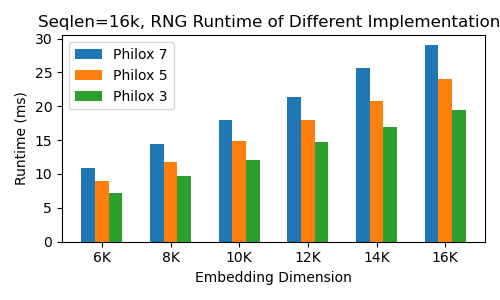}
\caption{Stand-alone RNG kernel runtime with different Philox implementations for a sequence length of 16K, measured in silicon. Less iterations lead to shorter runtime.}
\label{fig:seq16k_heads_v_runtime_philox357}
\end{figure}

\subsection{Implication of Cheaper RNG}
\label{sec:diff_rng}
In our prior discussions, we focused on the use of Philox 7 for RNG implementation. This subsection explores the implications of adopting more cost-effective RNG implementations, namely Philox 5 and Philox 3, which have fewer computational iterations and shorter runtimes.

We implemented all three Philox variants (Philox 3, 5, 7) on silicon using the GH100 with the same experimental setup as discussed. We show representative results collected for sequence length = 16K in Figure~\ref{fig:seq16k_heads_v_runtime_philox357}, which has a consistent trend with other configurations not shown on the graph.

We observed that the runtime of the Philox 5 RNG kernel is approximately 81\% of that required for Philox 7, while Philox 3 operates at 67\% of the Philox 7 runtime. These numbers align closely with the expected reduction in Fused Multiply-Add (FMA) operations (71\% for Philox 5 and 43\% for Philox 3). The difference is because other operations in the RNG process do not scale linearly with the number of iterations. 

Our analytical performance model was validated against the silicon results on overlapping RNG with QKV\_GEMM, showing an error margin consistent with previous validations. Using the validated model, we further analyze the implications of varying RNG complexities on overlapping RNG with all four GEMM kernels in the Transformer Block.

\begin{figure}[t]
\centering
\includegraphics[width=0.48\textwidth]{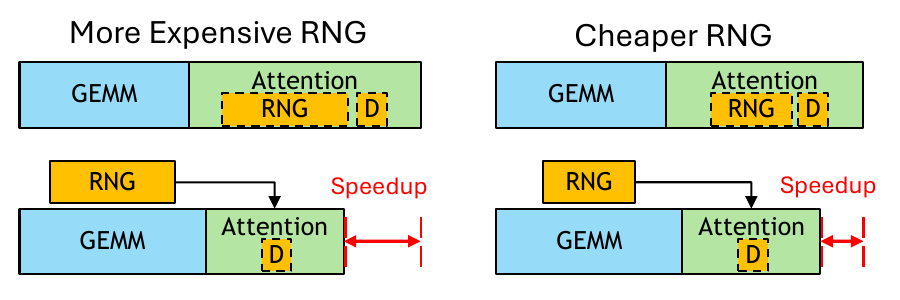}
\caption{Cheaper RNG implementation results in smaller overall speedup.}
\label{fig:reduced_rng_impact_to_overall}
\end{figure}

\begin{figure}[t]
\centering
\includegraphics[width=0.48\textwidth]{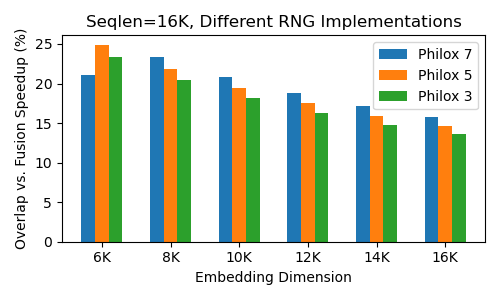}
\caption{Overlap speedup with different RNG Philox implementations, derived from theoretical model.}
\label{fig:seq16k_heads_v_runtime_philox357_speedup}
\end{figure}

\begin{figure*}[t]
\centering
\includegraphics[width=0.99\textwidth]{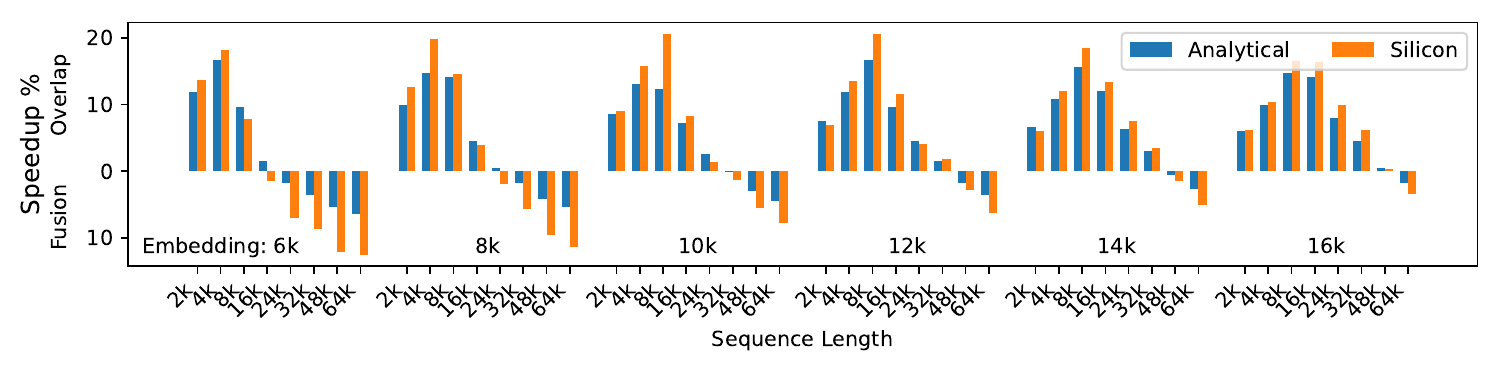}
\caption{Validation of our analytical model with silicon measurements, showing relative speedups of overlap and fusion methodologies. Data above the 0\% horizontal line indicates overlap's speedup over fusion, and below the line indicates fusion's speedup over overlap. }
\label{fig:validation}
\end{figure*}

As the complexity of the RNG algorithm decreases, the potential for speedup through overlapping also diminishes because the runtimes of the Attention and GEMM kernels remain unchanged. Thus, cheaper RNG implementations should result in smaller overall savings, as shown in Figure~\ref{fig:reduced_rng_impact_to_overall}. This is the case in most scenarios, as shown in Figure~\ref{fig:seq16k_heads_v_runtime_philox357_speedup}. However, in certain cases (such as Philox 7 implementation with embedding dimension = 6K and sequence length = 16K) the RNG runtime exceeds the GEMM runtime, introducing performance loss from fully exposed RNG after GEMM completes. This typically occurs with a smaller embedding dimension and a relatively long sequence length - Region 3 as discussed in Figure~\ref{fig:4gemm_result_speedup}. 

Moreover, we observed that the differences in speedup among various RNG implementations are relatively small. The standalone RNG kernel is primarily limited by the ALU pipeline, and the runtime decreases almost proportionally with the reduction in computation required. When RNG is fused into the Attention kernel, ALU is no longer the main bottleneck, and the performance depends on the Issue Stage where there is less difference between different RNG implementations. Consequently, the runtime reduction of the fused kernel is smaller than that of the standalone RNG kernel. Since the runtime of the GEMM kernel remains constant and the fused Attention-RNG kernel sees minimal changes, the overall speedup differences between the RNG implementations are relatively small.

\section{Validation with Silicon Implementation}
\label{sec:validation}

We validated the theoretical model's accuracy with a CUDA implementation on a NVIDIA H100 GPU~\cite{h100_whitepaper}, specifically using the GH100 HBM3 80GB variant. This GPU supports the latest High Bandwidth Memory (HBM) and provides ample compute resources and memory capacity for LLM training. Across multiple sequence length and embedding dimension configurations, we observed only 2\% average difference and 0.04 standard deviation between the analytical model and silicon measurements. The difference is calculated by averaging the absolute difference of speedup between silicon and theoretical results. 

\subsection{Implementation Details}

For GEMM kernel analysis, we implemented QKV\_GEMM, the GEMM layer that immediately precedes the Attention layer. Since GEMM layers have predictable runtime given the M, N and K dimensions, analyzing a single GEMM layer provides sufficient data to predict behavior across all four potential GEMM layers to overlap within a Transformer Block.

Our silicon implementation uses production-ready, highly optimized GEMM, Attention, and dropout kernels. 
The dropout kernel was modified to run in two scenarios: 1) fused within the Attention kernel, and 2) as a stand-alone RNG kernel storing bits representing random numbers in HBM for later use by the Attention kernel. We used the Philox 7 algorithm for RNG.

In terms of kernel scheduling, RNG and GEMM can either run as separate kernels or within the same kernel using warp specialization~\cite{li2023thread}\cite{warp_specialization}\cite{cutlass}. We opted for separate kernels to maintain a clear distinction between independent components in the network such as RNG and GEMM in this case. This approach allows other overlapping strategies to be implemented without deep knowledge of the kernels’ internal complexities, especially for the highly optimized GEMM kernels. Warp specialization enables a more fine-grained overlapping and we anticipate it to bring further, but limited, performance benefits. 

The baseline implementation in our experiments includes a QKV\_GEMM kernel, followed by a RNG kernel and an Attention kernel (including dropout), all run sequentially.

The fusion implementation includes a stand-alone kernel for QKV\_GEMM, and a kernel with fused dropout (including RNG) and Attention, running sequentially on the same CUDA stream. Our optimized kernel fusion minimizes synchronization overhead and maximizes the overlap of RNG operations with Attention's floating-point computations.

The overlapping implementation uses two separate CUDA streams to allow GEMM and RNG kernels to run concurrently. The Attention kernel is launched on the GEMM stream once both kernels have completed. This implementation minimizes the load latency of RNG states in Attention by overlapping the loads with the $Q*K$ matrix multiplication.

The GPU is warmed up using multiple instances of the same kernel sequence before actual measurements are taken. We validated our CUDA implementation against CPU generated result, ensuring correctness across various batch sizes, sequence lengths, and embedding dimensions.

\subsection{Validation Results}
Figure~\ref{fig:validation} shows the comparison of our analytical model and the silicon results. For better illustration purposes, instead of absolute runtime we show relative performance differences between RNG-Attention fusion and RNG-GEMM overlapping to validate our model's effectiveness for choosing the right technique for each configuration. The analytical model correctly derives the trade-off between the two options. Note that the results obtained here is different from what was shown in Figure~\ref{fig:4gemm_result_speedup}, because we only overlap with one GEMM instead of four GEMMs in our silicon implementation, simply for validation purposes. 

To further validate our assumptions about kernel interference and overhead performance impacts, we conducted the following tests:
\begin{itemize}
    \item \textbf{GEMM Resource Allocation:} Typically, the GEMM kernel utilizes all available Register Files and Shared Memory in each SM. Since RNG doesn't require much resources, in our implementation we carved out 6\% of the Registers and 7\% of the Shared Memory for the RNG kernel, hypothesizing this would not adversely affect GEMM performance. Our silicon measurements confirmed only 0.5\% average performance difference across various GEMM workload sizes.
    \item \textbf{Processing Time for Dropping Elements}: We hypothesized that dropping the elements within the Attention layer would require minimal runtime compared to RNG. Our silicon results confirmed that RNG dominated the full dropout, while dropping the elements only increase the original Attention runtime by 12\% on average. 
    \item \textbf{RNG and GEMM Interference:} We hypothesized that RNG should not noticeably slow down GEMM performance. We observed an average of 4\% slowdown in GEMM when running concurrently with RNG, which is acceptable. Conversely, RNG experiences a 50\% slowdown when run alongside GEMM, but this is also deemed acceptable since the original runtime of RNG is likely shorter than GEMM.
\end{itemize}
\section{Discussion}
\label{sec:discussion}

\subsection{Multi-GPU System}
We now extend our previous analysis on single GPU to a multi-GPU system, which is widely used in LLM training since the workload size is huge. The discussed insights for RNG-GEMM overlapping still hold true. 

Common strategies for running the workloads in parallel on multi-GPU include Tensor Parallelism~\cite{shoeybi2019megatron} and Sequence Parallelism~\cite{korthikanti2023reducing}, which split the head and sequence length dimensions, respectively. As depicted in Figure~\ref{fig:llm_network}, the multi-GPU setting introduces communication layers within each Transformer Block. The communication layers do not degradate the RNG-GEMM overlapping benefits: the latency of these communication layers can be optimized away by overlapping them with the GEMM layers' computation. Since communication layers utilize distinctly separate resources from GEMM (such as NVLink bandwidth), their concurrent execution will not introduce resource contention and will effectively hide communication latency. 

It is also worth noting that parallelism does not change the performance benefit ratios from overlapping RNG and GEMM analyzed earlier. Since the workload is split evenly on each GPU, the ratios of RNG, GEMM and Attention kernel runtime stay the same. 

\begin{figure}[t]
\centering
\includegraphics[width=0.48\textwidth]{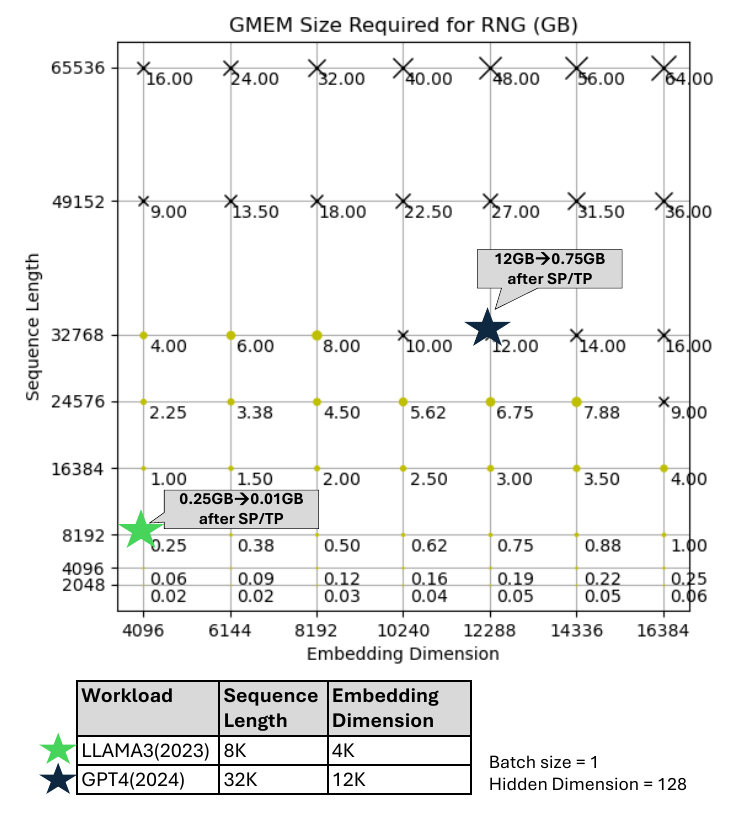}
\caption{Total memory capacity requirements for stand-alone RNG across different workload configurations.}
\label{fig:4gemm_result_hbm_size}
\end{figure}

\subsection{Memory Capacity Requirement}
Our overlapping methodology requires executing RNG as a stand-alone kernel which stores the generated RNG bits in memory to be later used by the Attention kernel. This is the same amount of memory usage as the baseline, sequential kernel implementation. For each element in Attention's intermediate matrix, assuming RNG generates 1 bit per element to indicate the dropping status, the entire RNG kernel requires storage of $B*nH*SQ^2$ in memory. Figure~\ref{fig:4gemm_result_hbm_size} illustrates the total memory capacity requirements for stand-alone RNG. 

\begin{figure}[t]
\centering
\includegraphics[width=0.48\textwidth]{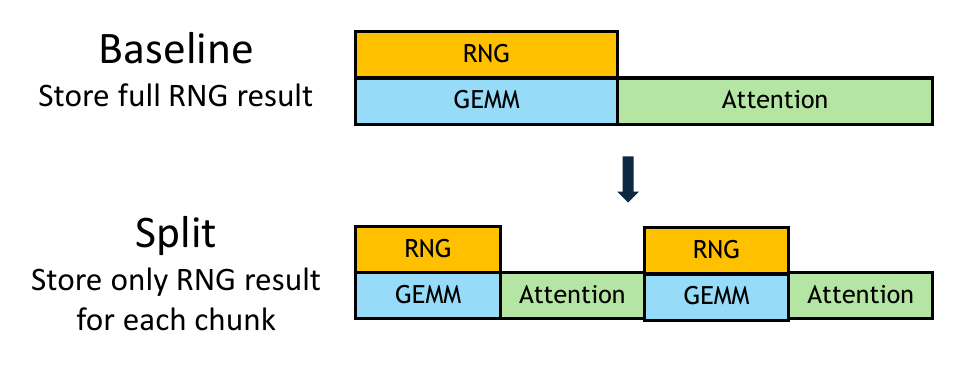}
\caption{Pipelining kernels that reduce memory storage requirements to negligible amount.}
\label{fig:pipelining_kernel}
\end{figure}

Since typical deployment scenario for LLM training involves multiple GPUs, the workload is divided into smaller segments distributed across the GPUs. This division significantly reduces the HBM capacity needed for RNG on each GPU. Figure~\ref{fig:4gemm_result_hbm_size} demonstrates the reduced HBM requirements with parallelism mechanisms. For the mainstream LLM training networks, the required HBM capacity is decreased by a factor of the number of GPUs used, depending on the chosen parallelism strategy and dimension (16 GPUs as shown in the figure).

To further alleviate memory storage concerns, an additional strategy can be applied which involves pipelining the RNG, GEMM, and Attention kernels. Figure~\ref{fig:pipelining_kernel} illustrates this strategy. It involves processing only a portion of the total computation per kernel at a time, splitting along the sequence length dimension to avoid creating dependencies in the GEMM kernel. With this methodology, we can reduce the RNG-GEMM overlapping's memory capacity requirements to arbitrarily small segments.

\section{Related Work}
Scheduling layer components within Large Language Models (LLMs) has become a critical area of research due to the increasing demands of improving the efficiency of these models~\cite{ye2024deepsurvey1}~\cite{li2024llmsurvey2}. Effective scheduling can improve end-to-end network performance when components underutilize GPU resources, if they have distinct resource utilization patterns.

Several studies have explored different aspects of LLM scheduling. For instance, Splitwise~\cite{patel2024splitwise} proposes a technique to separate the prefill and decoding phases of LLM inference onto different machines because of their unique characteristics. Similarly, Muxserve~\cite{duanmuxserve} introduces a spatial-temporal multiplexing system that can flexibly co-locate or separate these phases to maximize runtime efficiency.

A popular scheduling approach involves overlapping computation intensive components, such as matrix multiplications (matmuls) that utilize floating-point (FP) units, with inter-GPU communication tasks. The large sizes of LLM networks often require the workload to be divided and executed in parallel on multiple GPUs, where communication layers are required. Several forms of parallelism have been proposed to enhance efficient model training and serving, including Data Parallelism~\cite{dp}, Tensor Parallelism~\cite{shoeybi2019megatron}, Expert Parallelism~\cite{ep}\cite{megatron_blog}, and Sequence Parallelism~\cite{sp}. Since there is dependency between data on different GPUs, parallelism involves essential communication layers that transfer large amount of data between each GPU. Given the substantial communication overhead and distinct resource requirements between the data-transfer-heavy communication layers and the computation-heavy GEMM layers, it is ideal to overlap them to improve runtime. Best practices suggest splitting these components into finer-grained chunks and pipelining them for efficient overlapping~\cite{pati2024t3}\cite{comm_overlap}. This method is often enhanced by fine-grained kernel fusion techniques to further optimize performance~\cite{chang2024flux}\cite{punniyamurthy2023optimizing}.

In addition, several theoretical frameworks have been developed to model and analyze LLM workload performance by pinpointing hardware bottlenecks. Examples include roofline models~\cite{llm_roofline}, large-scale simulation framework~\cite{agrawal2024vidur}\cite{agrawal2024metron}, and light-weight performance modeling approaches~\cite{zhang2024llmcompass}. Our work uses a similar approach of analyzing LLM training performance based on hardware constraints, and extends these methodologies by providing detailed insights into finer-grained architecture details to evaluate the effectiveness of layer fusion and overlapping strategies.
\section{Conclusion}

This paper proposed a strategic overlapping of RNG with GEMM layers to improve end-to-end LLM training efficiency. By decoupling RNG from the Dropout process and running it in parallel with the computationally intense GEMM operations, we effectively reduce the latency impact typically associated with RNG if fused with the Attention layer. Our approach optimizes the use of hardware resources by exploiting the distinct hardware demands of RNG and GEMM, and achieves a notable improvement in end-to-end training performance, with 1.22x speedup within a Transformer Block for Llama3 and similar improvements across various LLM architectures.

We develop a fine-grained analytical performance model, validated with silicon results, to provide deeper insights into overlapping different kernels in the LLM network and highlight potential areas for further improvements. The principles established here can extend to optimizing other network layers, offering a generalized strategy to analyze the implications and enhance the performance of future LLM systems. As LLMs continue to scale and the demand for computational efficiency grows, the theoretical model can serve as a valuable framework for evaluating future overlapping strategies maximize resource utilization and minimize training time.

\bibliographystyle{ACM-Reference-Format}
\bibliography{reference}

\end{document}